\documentclass[12pt]{iopart}
\usepackage{graphicx}
\usepackage{amssymb} 
\usepackage{amsthm}
\usepackage{epsfig}

\begin{document}

\title[Elliptic and Hexadecapole flow of charged hadron in viscous hydrodynamics]{Elliptic and Hexadecapole flow of charged hadron in viscous hydrodynamics
with Glauber and Color Glass Condensate  initial conditions for Pb-Pb collision at $\sqrt{s_{NN}}$=2.76 TeV}
\author{Victor Roy$^1$, Bedangadas Mohanty$^1$, and A.K. Chaudhuri$^2$}
\address{$^1$School of Physical Sciences, National Institute of Science Education and Research, Bhubaneswar-751005, INDIA}
\address{$^2$Variable Energy Cyclotron Centre, 1/AF, Bidhan Nagar, Kolkata-700064, INDIA}
\ead{victor@niser.ac.in}

\begin{abstract} 
The experimentally measured elliptic ($v_{2}$) and hexadecapole ($v_{4}$) flow of charged particles as a function of transverse momentum ($p_{T}$)
at midrapidity in Pb-Pb collisions at $\sqrt{s_{\mathrm NN}}$ = 2.76 TeV are compared 
with the relativistic viscous hydrodynamic model simulations.
 The simulations are carried out for two different initial energy
density profiles obtained from (i) Glauber model, and (ii) Color Glass Condensate (CGC) model. Comparison to experimental data for 10-20\% to 40-50\% centrality, shows that a centrality dependent shear viscosity to
entropy density ($\eta/s$) ratio with values ranging between 0.0 to 0.12 is needed to explain the $v_{2}$ data
for simulations with the Glauber based initial condition. Whereas for the CGC based initial conditions a
slightly higher value of $\eta/s$ is preferred, around 0.08 to 0.16.
From the comparison of the $v_{4}$ simulated results to the corresponding experimental measurements we observe
that for the centralities 20-30\% to 40-50\% the $\eta/s$ values lies between 0.0 to 0.12 for both the
initial conditions studied.  The $\eta/s$ values obtained from our studies for Pb-Pb collisions at
$\sqrt{s_{\mathrm NN}}$ = 2.76 TeV are compared to other studies which uses both transport and
hydrodynamic approaches.

\end{abstract}
\submitto{\JPG}
\maketitle


\section{Introduction}

One of the interesting aspects of the high energy heavy-ion collisions at the Large Hadron Collider (LHC) is that,  a deconfined state of matter consisting of quarks and gluons, if produced in the collisions, exists for a
sufficiently long duration so as to enable the study of its transport properties.
The shear viscosity ($\eta$) is one of the transport coefficients that can be estimated
by comparing the experimental data to the theoretical calculations based on
either transport or hydrodynamic formulations. Particularly the experimental measurements of
the azimuthal anisotropy of produced particles are compared to corresponding theoretical simulations
to obtain an estimate of shear viscosity to entropy density ratio ($\eta/s$).
Where the azimuthal anisotropy measurements in the experiments are characterized in terms of
various coefficients of the Fourier expansion of the azimuthal distributions of produced particles
relative to the reaction plane. The second coefficient, $v_{2}$, is called as the elliptic flow
and is the most widely used observable to extract the $\eta/s$ value.  Such approaches have been successfully
applied to the experimental data for the Relativistic Heavy-ion collider (RHIC) energies. The extracted
$\eta/s$ values at RHIC are found to be lie between 1 - 5 $\times$ 1/4$\pi$\cite{Roy:2012jb,Niemi:2011ix,Shen:2010uy,Chaudhuri:2009uk,Bozek:2009dw,Schenke:2010rr,Romatschke:2007mq}.  $\eta/s$ = 1/4$\pi$
is the lower bound predicted by P. Kovtun et al.~\cite{Kovtun:2004de} (known as KSS limit) for strongly coupled quantum fluids using the string
theoretical approaches.

At the LHC energies for Pb-Pb collision at $\sqrt{s_{NN}}$ = 2.76 TeV one expects the initial temperature
attained in the collisions to be higher compared to that at RHIC for Au-Au collisions
at $\sqrt{s_{NN}}$ = 200 GeV.  Hence it would be interesting to find out the value of  $\eta/s$ for
the QCD matter produced at a higher temperature at LHC compared to RHIC. The estimates so far carried
out based on both transport theory~\cite{Xu:2011fe,Plumari:2011re} and
viscous hydrodynamic simulations~\cite{Bozek:2011wa,Luzum:2010ag,Schenke:2011tv,Song:2012tv}
have shown that the
values of  $\eta/s$ at LHC energies are similar to those obtained at RHIC. The hydrodynamic approaches
used included 2+1D simulations with smooth Glauber based initial conditions and 3+1D simulations with fluctuating
initial conditions. These approaches mostly concentrated on explaining the measured $v_{2}$ for certain collision centrality classes. One of the dominant source of uncertainties in such estimates
is  the lack of complete knowledge of the initial conditions in the heavy-ion collisions.
In this paper, we discuss results from a 2+1D smooth viscous hydrodynamic simulations with
two different initial conditions, one based on Glauber model and other based on CGC approach.
These simulations are then compared to centrality dependence of experimentally
measured $v_2$ as well as new data on hexadecapole flow ($v_{4}$) as a function of transverse
momentum from ATLAS~\cite{ATLAS:2011ah} and ALICE experiments~\cite{Abelev:2012di}
at LHC. 
A similar  work can be found in~\cite{Luzum:2010ae} , where the comparison of  viscous hydrodynamics 
simulation to the experimental data on $v_{2}$ and $v_{4}$ measured at top RHIC energy was
 done.
Actually in each collision the participant plane for both the second and fourth flow coefficient will fluctuate around the reaction plane.  The non-zero $v_{2}$ and $v_{4}$ in smooth hydrodynamics possibly can be interpreted as the average of many event-by-event hydrodynamics. It is then valid to compare the simulated $v_{2}$ and $v_{4}$ obtained for smooth initial condition in hydrodynamics with the experimental data. Our estimated values of $\eta/s$ based on simulations with two different initial conditions and two observables, $v_{2}$ and $v_{4}$ are then compared to other existing results in the literature.

The paper is organized as follows. In the next section we discuss the details of our 2+1D
viscous hydrodynamic simulations. These includes the initial conditions, an equation of state
based on the lattice QCD  and hadron resonance gas, and the freeze-out criteria. In section
3 we present the comparison of our simulation results to the corresponding experimental data
at midrapidity for Pb-Pb collisions at $\sqrt{s_{NN}}$ = 2.76 TeV. These includes the centrality
dependence of the charged particle pseudorapidity density ($dN_{ch}/d\eta$), $v_{2}(p_{T})$,
and $v_{4}(p_{T})$. In section 4 we present the predictions for charged hadron $p_{T}$
spectra from our simulations for various collision centralities in Pb-Pb collisions at
$\sqrt{s_{NN}}$ = 2.76 TeV. In section 5 we compare our estimated $\eta/s$
values to other calculations. Finally we summarize our work and present a brief outlook
in the section 6.

\section{2+1D viscous hydrodynamic simulations}

The hydrodynamic simulation results presented here are based on the second order causal Israel-Stewart
formalism for the evolution of a viscous fluid using the 2+1D viscous hydrodynamic
code ``AZHYDRO-KOLKATA''~\cite{Chaudhuri:2008sj,Roy:2011pk}. Shear viscosity is the only
dissipative process considered in our present study. We assume a net-baryon free plasma
is formed in Pb-Pb collisions at midrapidity at $\sqrt{s_{NN}}$ = 2.76 TeV, which
is supported by the value of anti-proton to proton ratio being close to unity~\cite{Aamodt:2010dx}.
The code solves the following  energy-momentum conservation equation along with the relaxation
equation for shear viscosity.
\begin{eqnarray}
\partial_\mu T^{\mu\nu} & = & 0,  \label{eq1} \\
D\pi^{\mu\nu} & = & -\frac{1}{\tau_\pi} (\pi^{\mu\nu}-2\eta \nabla^{<\mu} u^{\nu>}) \nonumber \\
&-&[u^\mu\pi^{\nu\lambda}+u^\nu\pi^{\mu\lambda}]Du_\lambda. \label{eq2}
\end{eqnarray}
Where $T^{\mu\nu}=(\varepsilon+p)u^\mu u^\nu - pg^{\mu\nu}+\pi^{\mu\nu}$ is the energy-momentum tensor,
$\varepsilon$, $p$, $u$, and  $\pi^{\mu\nu}$are the energy density, pressure, fluid
velocity and shear tensor respectively. $D=u^\mu \partial_\mu$ is the convective time derivative,
$\nabla^{<\mu} u^{\nu>}= \frac{1}{2}(\nabla^\mu u^\nu + \nabla^\nu u^\mu)-\frac{1}{3}                                                                      
(\partial_{\mu} u^{\mu}) (g^{\mu\nu}-u^\mu u^\nu)$ is a symmetric traceless tensor.
$\eta$ is the shear viscosity and $\tau_\pi$ is the corresponding relaxation time. Assuming boost-invariance, the equations are solved in ($\tau$,$x$,$y$,$\eta$) coordinate system.

\subsection{Initial conditions}

The initial conditions used here includes the initial energy density profile
in the transverse plane ($\epsilon (x,y)$). The $\epsilon (x,y)$ is obtained using two
different models - Glauber and CGC, the initial time
($\tau_{i}$), the transverse velocity profile $(v_{x}(x,y),v_{y}(x,y))$, shear stresses
in the transverse plane ($\pi^{\mu\nu}(x,y)$)  at $\tau_{i}$ are other input parameters. The
$\tau_{i}$ value is taken as 0.6 fm. The temperature independent $\eta/s$ values which are
used as an input to the viscous hydrodynamics simulation are considered to be 0, 0.08, 0.12, and 0.16. Generally the ratio $\eta(T)/s(T)$ is a function of temperature in both the QGP and hadronic
phases, the exact form of $\eta(T)/s(T)$ is still unknown~\cite{Niemi:2011ix,Song:2011qa,Shen:2011kn,Chaudhuri:2011np} . We decided to use constant $\eta(T)/s(T)$ value in our study, which may be thought of as a space-time average of the actual temperature dependent $\eta(T)/s(T)$.

The two component Glauber model based initial condition for the transverse energy
density is constructed as follows. At an impact
parameter $\bf{b}$ for Pb-Pb collisions at $\sqrt{s_{NN}}$ = 2.76 TeV, the transverse energy density
is obtained as,

\begin{equation}                                                                                                                                           
\epsilon({\bf b},x,y) = \epsilon_0[(1-x_{h})\frac{N_{part}}{2} ({\bf b},x,y)+ x_{h} N_{coll}({\bf b},x,y)]
\label{eq:enprof}
\end{equation}

\noindent where $N_{part}({\bf b},x,y)$ and $N_{coll}({\bf b},x,y)$ are the
transverse profile of participant numbers and binary collision numbers
respectively. The $\epsilon_0$ corresponds to the central energy density for $b$ = 0 collision and does
not depend on the impact parameter of the collision. The parameter $x_{h}$ is
the hard scattering fraction.  The value of $x_{h}$ is fixed by reproducing the experimentally measured centrality dependence of charged hadron multiplicity density at midrapidity~\cite{Roy:2011xt}.
For each value of $\eta/s$ the magnitude of $\epsilon_{0}$ is adjusted to reproduce the experimental charged hadron multiplicity density  for the central 0-5\% collisions at midrapidity ($dN_{ch}/d\eta$) which is 1601 $\pm$ 60~\cite{Aamodt:2010cz}.
No further adjustment of the parameters are
done for other collision centralities. The values of $\epsilon_{0}$ for different
values of $\eta/s$ are shown in Table~\ref{table1}.
The $N_{part}({\bf b},x,y)$ and $N_{coll}({\bf b},x,y)$
values are obtained using an optical Glauber model calculation~\cite{Miller:2007ri}.
The inelastic nucleon-nucleon cross section is taken as 64 mb~\cite{Aamodt:2010cz}.

\begin{table}[h]
\begin{center}
\caption{\label{table1} Values of $\epsilon_{0}$ used in Glauber, CGC model and the normalization constant $C$ used in CGC model for initial transverse
energy density.}
\begin{tabular}{|c|c|c|c|}
\hline
$\eta/s$ &  $\epsilon_0$ (GeV/$fm^{3}$), Glauber & $C$ (GeV/$fm^{1/3}$) CGC  & $\epsilon_0$(GeV/$fm^{3}$), CGC   \\\hline
0.0          &   112         & 0.033    & 77.7     \\\hline
0.08         &   98          & 0.030    &70.6    \\ \hline
0.12         &   88          & 0.027    & 63.5    \\ \hline
0.16         &   78          & 0.024    & 56.5     \\ \hline
\end{tabular}
\end{center}
\end{table}

The CGC based initial transverse energy density profile is obtained as
follows. We follow references \cite{Dumitru:2007qr,Luzum:2008cw} and consider that the initial
energy density can be obtained from the gluon number density through
the thermodynamic relation,
\begin{equation}
 \epsilon(\tau_i,{\bf x}_T,b)={\rm C}\times \left[ \frac{dN_g}{d^2 {\bf x}_{T}dY}({\bf x}_T,b)\right] ^{4/3},
\label{edCGC}
\end{equation}
where $\frac{dN_g}{d^2 {\bf x}_{T}dY}$ is the gluon number density evaluated at
central rapidity $Y=0$ and  the overall normalization $C$ is a free
parameter. $C$ is fixed to reproduce the experimentally
measured charged particle multiplicity density at midrapidity. The
values of $C$ used in the simulations for different
input values of $\eta/s$ and the corresponding central energy densities are given in Table~\ref{table1}. The value of $\epsilon_{0}$ given in table~\ref{table1} for CGC initial condition 
corresponds to the central energy density for collision with impact parameter b=2.21 fm (0\%-5\% centrality).
The details of the implementation of this approach can be found in Refs~\cite{Roy:2012jb,Kharzeev:2002ei,Drescher:2006pi}.

Finally the shear stresses $\pi^{\mu\nu}$ are initialized to their corresponding
Navier-Stokes estimates for the boost invariant velocity profile,
$\pi^{xx}=\pi^{yy}=2\eta/3\tau_i$, and $\pi^{xy}=0$ \cite{Chaudhuri:2008je}.
We have used $\tau_\pi=3\eta/4p$ (where $\eta$ and $p$ are the shear viscous coefficient
and pressure) in our simulation, which corresponds to the
kinetic theory estimates of relaxation time for shear viscous stress for a
relativistic Boltzmann gas \cite{Israel:1979wp}. The initial values of
$v_{x}(x,y)$ and $v_{y}(x,y)$ are taken to be zero.

\subsection{Equation of state}

The equation of state used here is a combination of high temperature phase from
the Lattice QCD result with cross-over transition at temperature $T_c$ = 175 MeV
 \cite{Borsanyi:2010cj} and the low temperature phase is modeled by hadronic resonance gas,
containing all the resonances with $M_{res} \leq$2.5 GeV~\cite{Roy:2011xt}.
Entropy density ($s$) of the two phases are joined at $T$ = $T_c$ = 175 MeV by a smooth step
like function.  The thermodynamic variables pressure ($p$), energy
density ($\varepsilon$), etc. are  then calculated by using the standard
thermodynamic relations.
\begin{eqnarray}
p\left(T\right)&=&\int^{T}_{0}s\left(T^{\prime}\right)dT^{\prime}\\
 \varepsilon\left(T\right)&=&Ts\left(T\right)-p\left(T\right).
\end{eqnarray}

\subsection{Freeze-out condition}

We use the Cooper-Frye algorithm at the freezeout to calculate
invariant yields of the hadrons~\cite{Cooper:1974mv} .
The freezeout  temperature which is a free parameter in the
hydrodynamics simulation is taken as
$T_{f}$=130 MeV.

The freezeout distribution function in presence of viscous effects gets modified
and can be approximated as \cite{Chaudhuri:2008sj}
\begin{eqnarray}
f_{neq}(x,p)=f_{eq}(x,p)[1+\phi(x,p)],
\end{eqnarray}
where $\phi(x,p)<<1$ is the corresponding deviation from the equilibrium
distribution function $f_{eq}(x,p)$. The non-equilibrium correction
$\phi(x,p)$ can be calculated in Grad's 14 moment method. According to which the viscous
correction terms up to a quadratic function
of the four momentum $p^{\mu}$ \cite{Muronga:2004sf,Muronga:2006zx}.
As only shear stresses are considered, $\phi(x,p)$ has the following form
\begin{eqnarray}
\phi(x,p)=\varepsilon_{\mu\nu}p^{\mu}p^{\nu},
\end{eqnarray}
where
\begin{eqnarray}
\varepsilon_{\mu\nu}=\frac{1}{2(\epsilon+p)T^{2}}\pi_{\mu\nu}.
\end{eqnarray}
The correction factor thus increases with increasing values
of shear stress $\pi_{\mu\nu}$ at freezeout. The correction term also
depends on the particle momentum. Though we have shown comparison of 
viscous hydrodynamics simulation to the experimental data up to $p_{T}$=4 GeV,
the applicability of the formalism is not reliable beyond $p_{T}\approx$ 3 GeV because of the momentum dependence of viscous correction. 
The Cooper-Frye formula \cite{Cooper:1974mv} for a non equilibrium system is
\begin{eqnarray}\nonumber
\frac{dN}{d^{2}p_{T}dy}&=&\frac{g}{(2\pi)^{3}}\int d\Sigma_{\mu}p^{\mu}f_{neq}(p^{\mu}u_{\mu},T), \\\nonumber
\label{eq:chap3_shear_cooper}
\end{eqnarray}
where $g$ is the degeneracy of the particle considered and
$d\Sigma_{\mu}$ is the normal to the elemental freeze-out hypersurface.

\section{Comparison to experimental data}

The experimental data used for comparison to the simulated results
are from the ALICE and ATLAS experiments at LHC. The observables
are charged hadron multiplicity density ($dN_{ch}/d\eta$) at midrapidity~\cite{Aamodt:2010cz},
elliptic flow~\cite{ATLAS:2011ah},  and hexadecapole flow~\cite{Abelev:2012di} as a function of $p_{T}$ measured at
midrapidity for Pb-Pb collisions at $\sqrt{s_{NN}}$ = 2.76 TeV. All
results are for collision centrality 10-20\%, 20-30\%, 30-40\%, and
40-50\%. For charged hadron yields we also present results for 0-5\% collision centrality. This centrality is used to fix the initial energy density parameter in our simulations.

The $v_{2}(p_{T})$ data is from the  ATLAS experiment~\cite{ATLAS:2011ah}
and are obtained using the formula $v_{2} = \langle cos(2 (\phi - \Psi_{2})) \rangle$
after correction of the event plane resolution. Where $\phi$ is the
azimuthal angle of the charged hadrons measured within $\mid \eta \mid$ $<$ 1
and $\Psi_{2}$ is the second order event plane constructed using a
forward calorimeter detector in $3.2 < \eta < 4.9$. The rapidity gap between the detectors used to
measure the $v_{2}$ and  $\Psi_{2}$  ensures absence of significant
$\Delta \eta$ dependent non-flow correlations, which are also absent
in our hydrodynamic simulations. $\Psi_{2}$ is zero, i.e, the impact parameter vector 'b' is along $x$ axis.

The $v_{4}(p_{T})$ data is from the ALICE experiment~\cite{Abelev:2012di}
and are obtained using the formula $v_{4} (\Psi_{4})= \langle cos(4 (\phi - \Psi_{4})) \rangle$
and $v_{4} (\Psi_{2})= \langle cos(4 (\phi - \Psi_{2})) \rangle$
after correction of the event plane resolution. Where $\phi$ is the
azimuthal angle of the charged hadrons measured within $\mid \eta \mid$ $<$ 0.8
and $\Psi_{4}$ is the fourth order event plane constructed using  two scintillator arrays
of detectors covering $-3.7 < \eta < -1.7$ and $ 2.8 < \eta < 5.1$. Like
$v_{2}(p_{T})$, the rapidity gap between the detectors used to
measure the $v_{4}$ and  $\Psi_{4}$  ensures absence of significant
$\Delta \eta$ dependent non-flow correlations, which are also absent
in our hydrodynamic simulations. $\Psi_{4}$ is zero for our simulation and impact parameter 'b'  is along $x$ axis. Since we are measuring flow harmonics with respect to the second order
event plane $\Psi_{2}$, it will be more appropriate to compare our simulated result
of $v_4$ with the experimental data $v_{4} (\Psi_{2})$.

\begin{table}
\caption{\label{table2}Comparison of experimentally measured $dN_{ch}/d\eta$ at midrapidity for Pb-Pb collisions at  $\sqrt{s_{\rm {NN}}}$ = 2.76 TeV for various collision centrality~\cite{Aamodt:2010cz} to those obtained from a 2+1D viscous hydrodynamic simulations
with Glauber and CGC based initial conditions.}
\vspace{.5cm}
\begin{center}
\begin{tabular}{|c|cccccc|}
\hline
\% cross section &  &  & $dN_{ch}/d\eta$   &   &  &\\
      & Experiment           &            & Glauber  &   &   &  \\
      &           & $\eta/s$ = 0.0           & $\eta/s$ = 0.08   &   $\eta/s$ = 0.12  & $\eta/s$ = 0.16     &   \\
\hline
0-5       &1601 $\pm$60  &1569  & 1572 &1567 &1565&\\
10-20  & 966 $\pm$ 37  & 959    & 968  & 966  & 967 &\\
20-30  & 649 $\pm$ 23  & 637    & 647  & 649  & 651 &\\
30-40  & 426 $\pm$ 15  & 414    & 424  & 427  & 428 &\\
40-50  & 261 $\pm$ 9   & 248    & 256  & 260  & 260 &\\
\hline
\% cross section  & Experiment           &            & CGC  &   &  &   \\
      &           & $\eta/s$ = 0.0           & $\eta/s$ = 0.08   &   $\eta/s$ = 0.12  & $\eta/s$ = 0.16  &   \\
\hline
0-5       &1601 $\pm$60  & 1577  &1629 &1623 &1626 &\\
10-20  & 966 $\pm$ 37  & 974    & 981  & 985  & 988 &\\
20-30  & 649 $\pm$ 23  & 640    & 666  & 668  & 670 &\\
30-40  & 426 $\pm$ 15  & 424    & 443  & 446  & 447 &\\
40-50  & 261 $\pm$ 9   & 259    & 274  & 276  & 276 &\\
\hline
\end{tabular}
\end{center}
\end{table}

\subsection{Charged particle pseudorapidity density}

The initial energy density profiles are fixed in the simulations by matching the experimental measured
$dN_{ch}/d\eta$ for only 0-5\% central Pb-Pb collisions at  midrapidity for $\sqrt{s_{\rm {NN}}}$ = 2.76 TeV~\cite{Aamodt:2010cz}.
No adjustment of the parameters are done to match the $dN_{ch}/d\eta$ for the rest of the collision centralities
10-20\%, 20-30\%, 30-40\%, and 40-50\% studied in this paper. The table~\ref{table2} shows the comparison
between $dN_{ch}/d\eta$ at midrapidity as measured by ALICE for  Pb-Pb collisions at $\sqrt{s_{\rm {NN}}}$ = 2.76 TeV
to our simulation results for various values of $\eta/s$ for both Glauber and CGC based initial conditions.
We observe for the collision centrality classes studied, there is good agreement between the experimental
measurements and the simulated results for both the initial conditions.

\subsection{Elliptic flow}
\begin{figure}
\begin{center}
\includegraphics[scale=0.45]{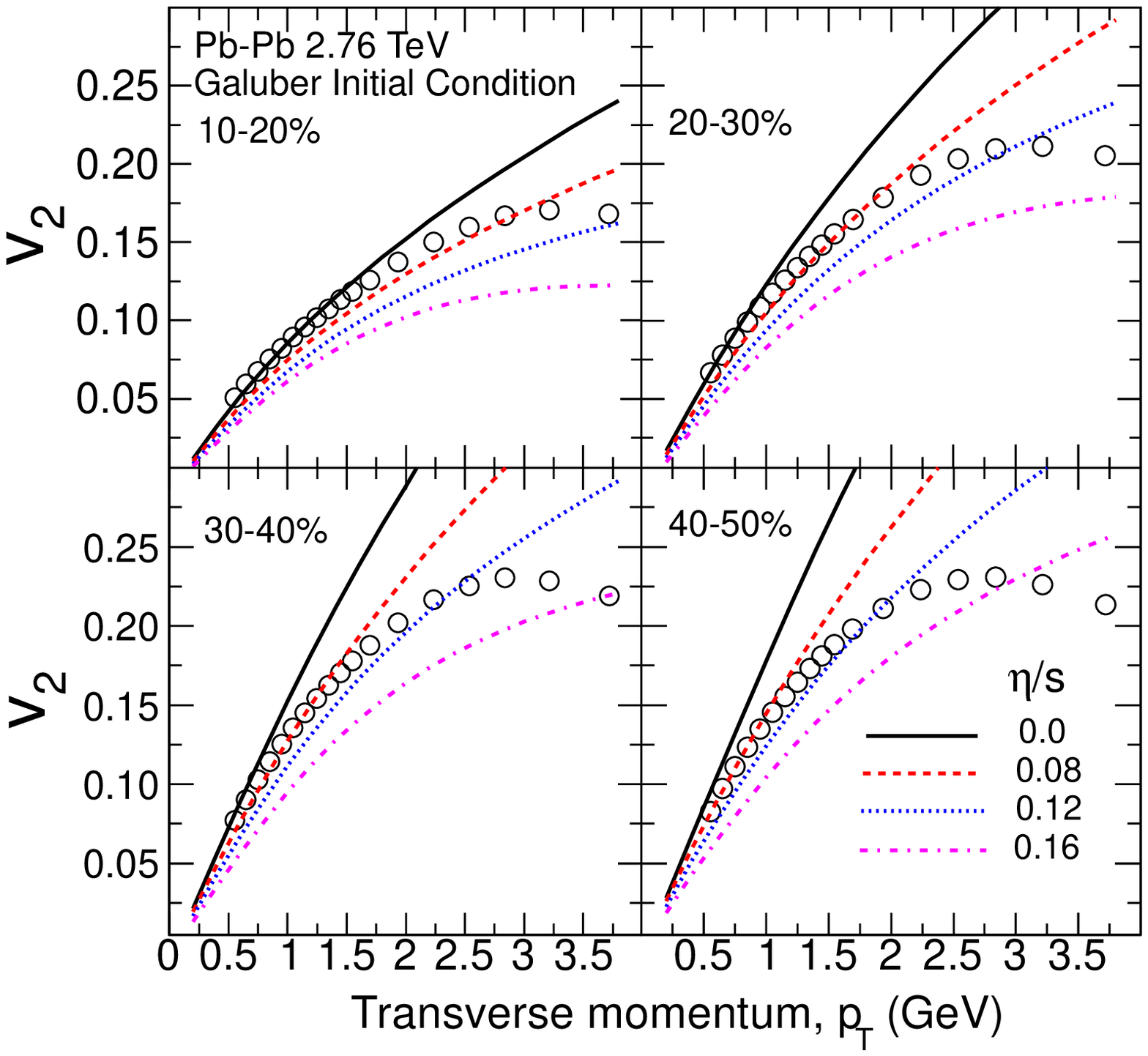}
\includegraphics[scale=0.45]{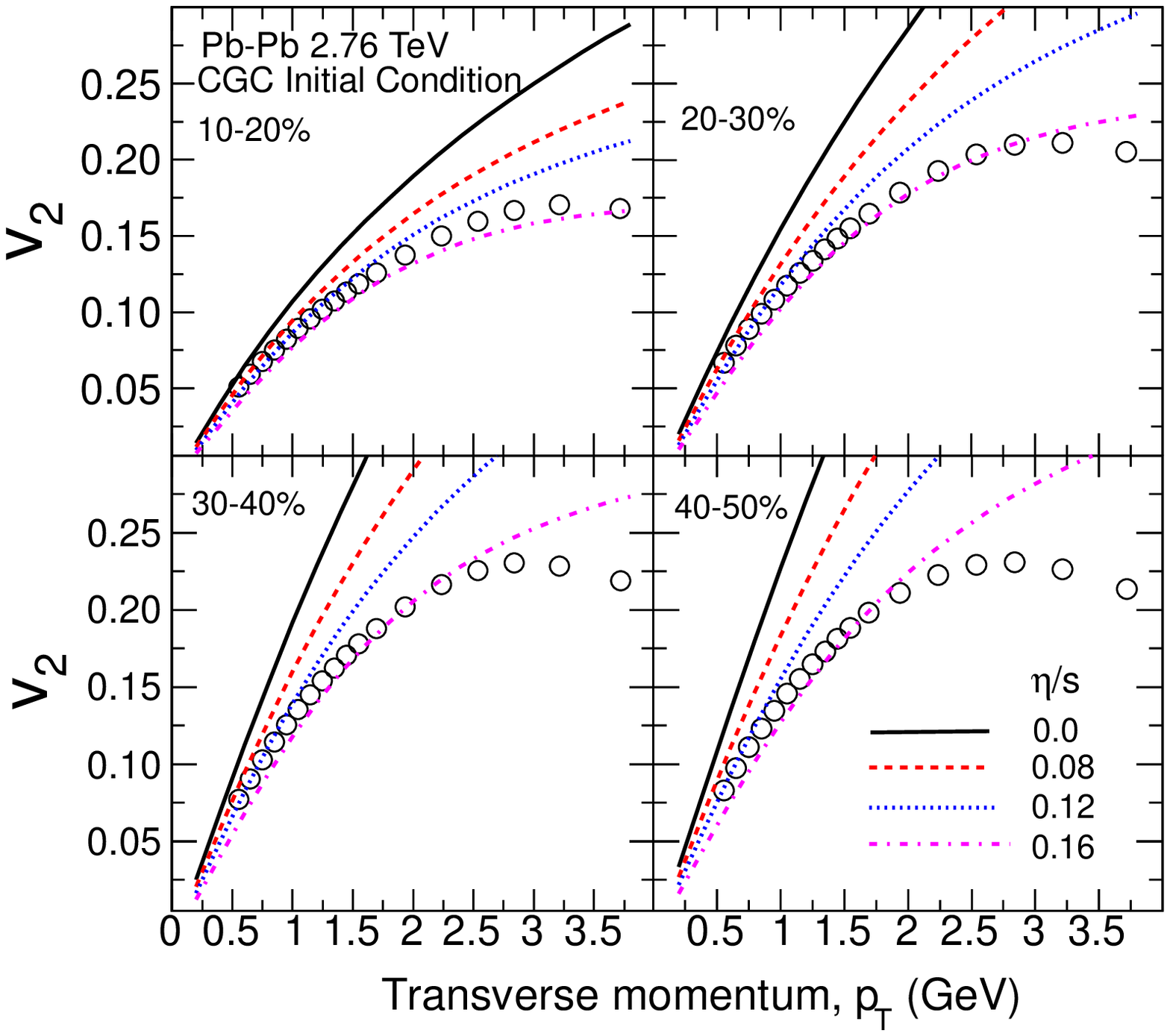}
\caption{(Color online) Elliptic flow of charged hadrons as a
  function of transverse momentum at midrapidity for Pb-Pb collisions
  at $\sqrt{s_{\rm {NN}}}$ = 2.76 TeV. The open circles corresponds to
  experimental data measured by the ATLAS collaboration~\cite{ATLAS:2011ah} . The lines
  represent results from a  2+1D relativistic viscous hydrodynamic model
  with a Glauber based (upper panel) and CGC based (lower panel)
  initial transverse energy density profile
  and different $\eta/s$ values.}
\label{fig1}
\end{center}
\end{figure}
Upper panel of Fig~\ref{fig1} shows the $v_{2}$ as a
function of $p_{T}$ for charged hadrons at
midrapidity ($\mid \eta \mid < 1$) for Pb-Pb collisions at
$\sqrt{s_{\rm {NN}}}$ = 2.76 TeV. The results are shown for four
different collision centralities (10-20\%, 20-30\%, 30-40\%, and
40-50\%). The open circles are the experimental data from the ATLAS
collaboration~\cite{ATLAS:2011ah}. The simulated results are from the 2+1D relativistic
viscous hydrodynamic model with a Glauber based initial transverse
energy density profile. The black solid, red dashed, blue dotted, and magenta
dot-dashed lines corresponds to calculations with
$\eta/s$ = 0.0, 0.08, 0.12, and 0.16 respectively. The experimental data for 10-20\% collision
centrality is best described by ideal fluid ($\eta/s$ = 0.0) simulation results
for most of the $p_{T}$ range for which comparison has been done.
The experimental data for 40-50\% collision centrality prefer a $\eta/s$ between 0.08 and  0.12.

Lower panel of Fig.~\ref{fig1} shows the same results as in the upper panel but now the simulated results corresponds to
2+1D viscous hydrodynamic calculations with a CGC based initial
transverse energy density profile. Comparison of the simulated results in the upper panel and lower panel
for the same collision centrality shows that the $v_{2}(p_{T})$ for CGC based initial condition
is larger compared to corresponding results from Glauber based initial
condition. This can be understood from the fact that CGC based
initial condition leads to a higher value of momentum anisotropy
compared to Glauber based initial condition~\cite{Roy:2012jb,Luzum:2008cw,Song:2010mg,
Hirano:2005xf}.
We find from the comparison of experimental data to
simulations  based on CGC initial conditions that the $v_{2}(p_{T})$
data for all the collision centralities studied are best explained for
with $\eta/s$ between 0.12-0.16. In contrast to Glauber based initial conditions
we find the CGC based initial conditions prefer a slightly higher value of $\eta/s$
for the LHC data.

\begin{figure}
\begin{center}
\includegraphics[scale=0.45]{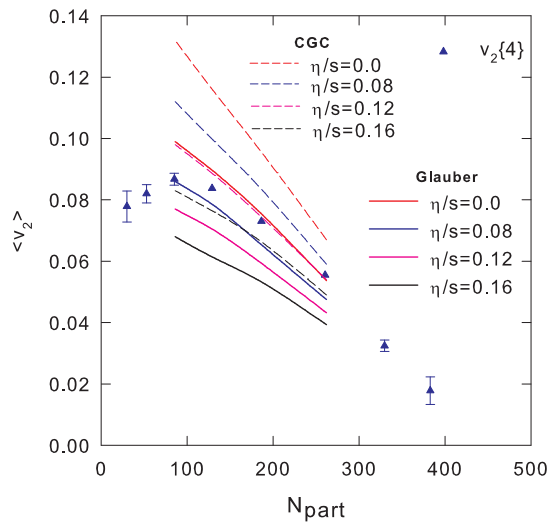}
\caption{(Color online) $p_{T}$ integrated elliptic flow of charged hadrons as a
  function of number of participant ($N_{part}$) at midrapidity for Pb-Pb collisions
  at $\sqrt{s_{\rm {NN}}}$ = 2.76 TeV. The experimental data from ALICE collaboration \cite{Aamodt:2010pa}  are shown by solid triangles. Viscous hydrodynamics simulation results for CGC (dashed lines) and Glauber model (solid lines) initial condition for $\eta/s$ =0 (red), 0.08 (blue),0.12 (pink), 0.16
  (black) are shown by lines. }
\label{fig1.2}
\end{center}
\end{figure}

In figure \ref{fig1.2} the comparison between experimentally measured $p_{T} $ integrated 
elliptic flow as a function of number of participant ($N_{part}$) and simulated results obtained from viscous hydrodynamics  for four different values of $\eta/s$  are shown by different lines.
The experimental data are from ALICE collaboration~\cite{Aamodt:2010pa} . The solid lines are simulated results
for Glauber initial condition and dashed lines are for CGC initial condition. The simulated integrated flow is shown here for the centrality class 10-50\%, as the experimental data on differential elliptic flow is available only for those centrality collisions and our simulation was
also done for the above range of collision centrality. Conclusion is consistent with those obtained
from differential elliptic flow $(v_2(p_T))$.

\subsection{Hexadecapole flow}
\begin{figure}
\begin{center}
\includegraphics[scale=0.5]{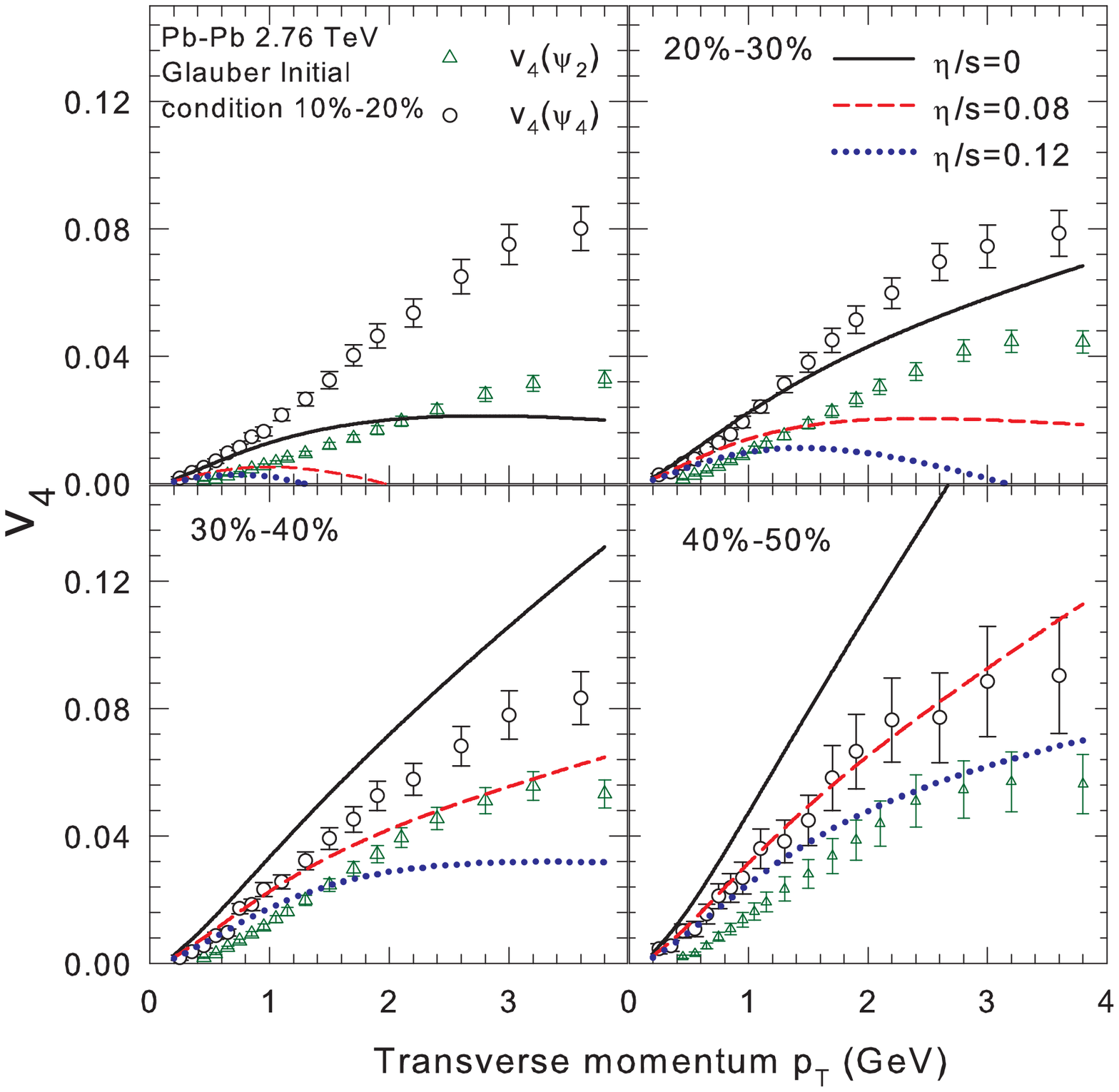}
\includegraphics[scale=0.5]{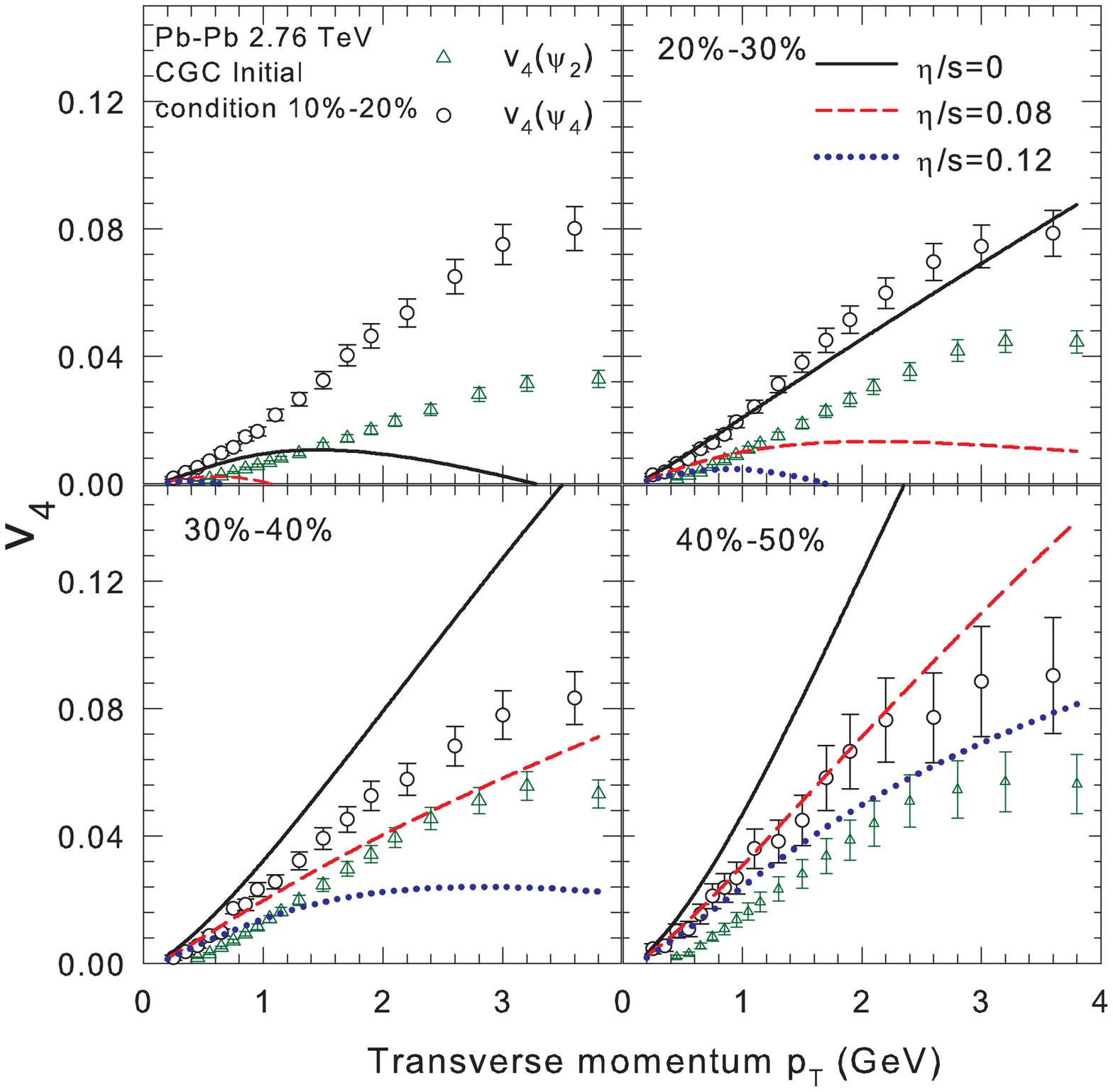}
\caption{(Color online) Hexadecapole flow of charged hadrons as a
  function of transverse momentum at midrapidity for Pb-Pb collisions
  at $\sqrt{s_{\rm {NN}}}$ = 2.76 TeV. The open circles corresponds to 
  the experimentally measured $v_{4}$ with respect to $\Psi_{4}$ and 
  open triangles corresponds to measured $v_{4}$ with respect to 
  $\Psi_{2}$ by the ALICE collaboration~\cite{Abelev:2012di}. 
  The lines represent results from a  2+1D relativistic viscous hydrodynamic 
  model with both Glauber based (upper panel) and CGC based  (lower panel) 
  initial transverse energy density profile for different $\eta/s$ values.
}
\label{fig2}
\end{center}
\end{figure}

Fig.~\ref{fig2} shows the hexadecapole flow ($v_{4}$) as a
function of transverse momentum ($p_{T}$) for charged hadrons at
midrapidity in Pb-Pb collisions at $\sqrt{s_{\rm {NN}}}$ = 2.76 TeV.
The results are shown for four different collision centralities (10-20\%, 20-30\%, 30-40\%, and
40-50\%). The open circles are the experimental data $v_{4}(\psi_{4})$ from the ALICE
collaboration~\cite{Abelev:2012di}, the model simulations are also compared to $v_{4}$ measured with respect to the second order event plane $\Psi_2$ by ALICE collaboration.

The upper panel shows the simulated results for ideal and viscous
fluid evolution using Glauber based initial condition,
while the simulated results for the CGC based initial conditions are shown
in the lower panel. 

 We find that $v_{4}(p_T )$ from ideal hydrodynamic simulation with Glauber based initial conditions underestimate the experimental $v_{4}$ data measured with respect to $\Psi_4$ for 10-20\% collision centrality and agrees with the experimental $v_{4}$ data measured with respect to $\Psi_4$ at lower $p_T$ for 20-30\% collision centrality. In more peripheral collisions 30-40\% and 40-50\% centrality, the comparison of the simulated results to the experimental v4 data with respect to $\Psi_4$ shows that the preferred $\eta/s$ lies between 0.0 and 0.08. A similar conclusion is also made from the comparison between simulated results with CGC based initial condition and experimental $v_{4}$ data with respect to $\Psi_4$.  From the comparison of the simulated results with respect to experimental $v_{4}$ data measured with respect to $\Psi_2$ we observe that at low $p_T$ ideal hydrodynamics over estimates the data for all centralities and for both the initial conditions studied.  As we move to peripheral collisions the preferred values of $\eta/s$ increases from 0.0 to 0.12 for collision centrality range 10-20\% to 40-50\% respectively for both the initial conditions.  We also observe that the higher order flow
coefficient $v_{4}(p_{T})$ is more sensitive than $v_{2}(p_{T})$ to the shear viscosity of the fluid. It has been shown in reference~\cite{Teaney:2012ke} and references therein that
in conformal viscous fluid and for a particular fluid flow profile each linearized perturbation 
labeled by n, m-th cumulant is damped by a factor $\sim$$ exp(-\Gamma_{n,m}\tau_{final})$ relative
to ideal hydrodynamics, where  $\tau_{final}$ is the lifetime of the fluid evolution. Where
$\Gamma_{n,m}\sim \frac{l_{mfp}}{(L \tau_{final})}\left(\frac{n-m}{2}+m\right)^{2},$ and
$l_{mfp}$, L are mean free path and the system size respectively. Then it was 
shown that viscous correction to first ($v_{1}$) and second order harmonic flow ($v_{2}$) is of
the same order, whereas for $v_{4}$ the correction will be four times more compared to $v_{2}$.
In accordance with the above reasoning we can understand the higher viscous correction for $v_{4}$ compared to $v_{2}$ for our case. A similar behavior was also observed for a viscous hydrodynamics simulation in reference~\cite{Schenke:2011bn,Alver:2010dn}.

We observe from the comparison of our simulated result on $v_2$ and $v_4$ that the fluid viscosity increases for peripheral collision compared to central collisions. This fact may be understood qualitatively in the following way~\cite{Chaudhuri:2009hj}. For the peripheral collision both the system size and initial central temperature of the reaction zone is smaller than the central collision. Consequently the lifetime of the fireball for peripheral collision is shorter in comparison to central collision. This lower initial central temperature and reduced lifetime implies that for peripheral collisions the fluid spend most of its lifetime in the hadronic phase. And it is known that the shear viscosity of hadronic phase is larger compared to QGP phase. Thus the effective viscosity (which we are measuring) of the fluid will be greater for peripheral collisions compared to central collision.

\section{Simulated charged hadron transverse momentum spectra}

\begin{figure}
\begin{center}
\includegraphics[scale=0.35]{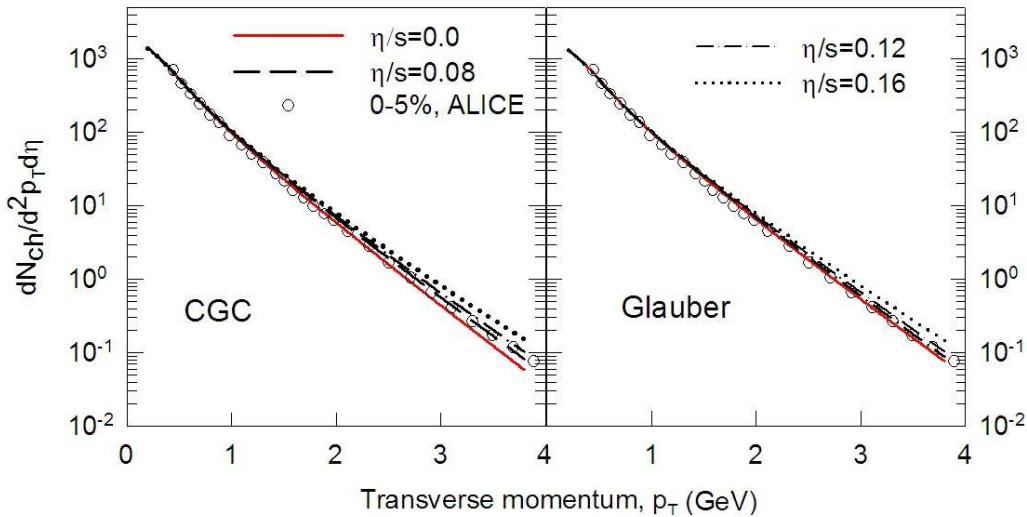}
\caption{(Color online) Invariant yield of charged hadrons as a
  function of transverse momentum at midrapidity for Pb-Pb collisions
  at $\sqrt{s_{\rm {NN}}}$ = 2.76 TeV. The open circles corresponds to
  experimental data for 0-5\% centrality collision measured by the ALICE 
  collaboration~\cite{Aamodt:2010jd} . The lines
  represent results from a  2+1D relativistic viscous hydrodynamic model
  with a CGC based (left panel) and Glauber based (right panel)
  initial transverse energy density profile
  for different $\eta/s$ values.}
\label{fig3a}
\end{center}
\end{figure}

We have compared our simulated results for charged hadron transverse momentum
spectra with the available experimental data for 0-5\% centrality Pb-Pb collision at LHC~\cite{Aamodt:2010jd}.
The comparison is shown in Fig~\ref{fig3a} . The left panel of Fig~\ref{fig3a} is for CGC initial 
condition whereas the right panel of the same figure shows the result obtained
for Glauber initial condition. We have fixed the initial energy density of the fluid and the freeze-out
temperature  by reproducing the experimental transverse momentum spectra of charged hadron
for 0-5\% collision centrality. So it will be illogical to extract viscosity by comparing simulation results to the experimentally measured transverse momentum spectra for 0-5\% centrality. 

\begin{figure}
\begin{center}
\includegraphics[scale=0.5]{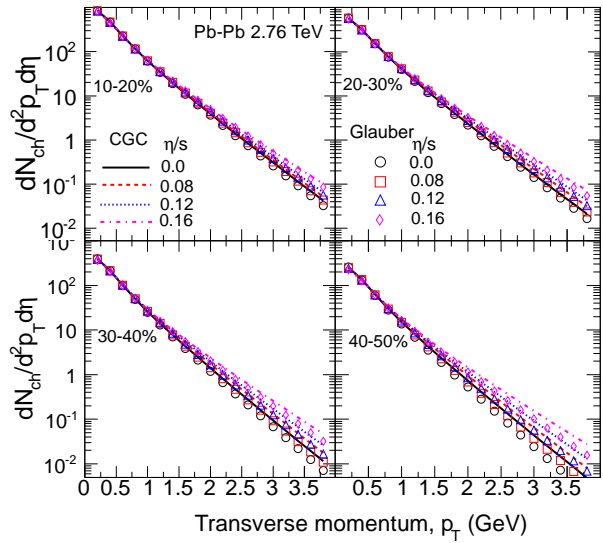}
\caption{(Color online) Invariant yield of charged hadrons as a
  function of transverse momentum at midrapidity for Pb-Pb collisions
  at $\sqrt{s_{\rm {NN}}}$ = 2.76 TeV from a  2+1D relativistic
viscous hydrodynamic model. The open symbols are for results obtained using
Glauber based initial transverse energy density profile and different $\eta/s$ values.
The lines corresponds to calculations with CGC based initial transverse energy density profile.
The results are shown for four different collision centralities 10-20\%,
20-30\%, 30-40\%, and 40-50\%.}
\label{fig3}
\end{center}
\end{figure}

Fig~\ref{fig3} shows the simulated invariant yield of charged
hadrons as a function of transverse momentum at midrapidity for Pb-Pb
collisions at $\sqrt{s_{\rm {NN}}}$ = 2.76 TeV for four different
collision centralities (10-20\%, 20-30\%, 30-40\%, and
40-50\%). The open symbols corresponds to the simulated results
from the 2+1D relativistic viscous hydrodynamic model with a Glauber
based initial transverse energy density profile while the lines
corresponds to the calculations with a CGC based initial condition.
The black open circle (black solid line), red open square (red dashed line),
blue open triangle (blue dotted line) and magenta diamond (magenta dash-dotted line)
symbols corresponds to simulations with $\eta/s$ = 0.0, 0.08, 0.12, and 0.16
respectively. As we go from 10-20\% collision centrality to 40-50\% collision
centrality for CGC based initial condition, spectra become more flatter than
for Glauber initial condition.  This means that the
average transverse velocity at the freeze-out which determines the
slope of the $p_{T}$ spectra is slightly larger for the fluid
evolution with CGC than those corresponding to the Glauber based initial
conditions.

\section{Comparison of our estimate of $\eta/s$ with other calculations}
\begin{figure}
\begin{center}
\includegraphics[scale=0.4]{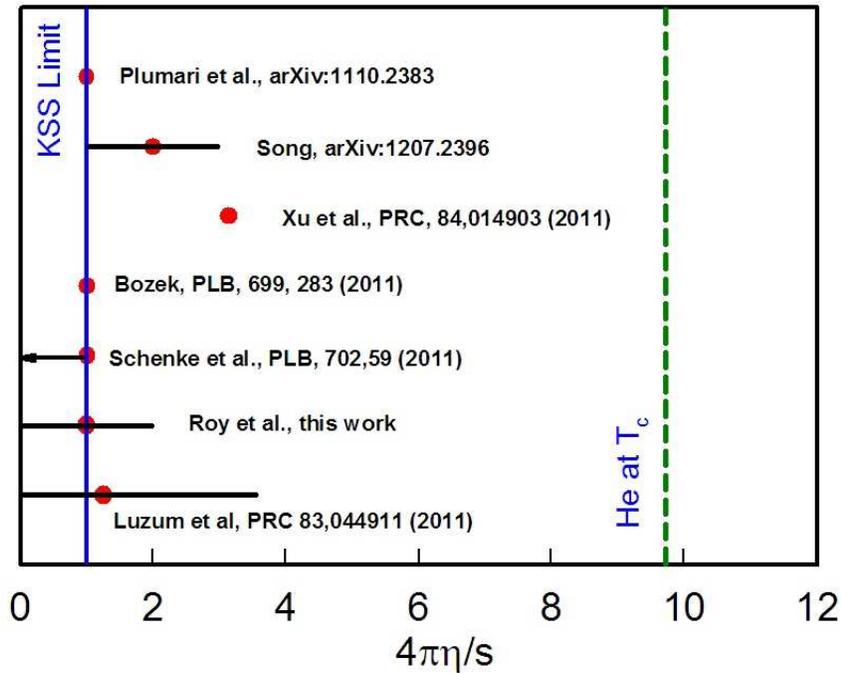}
\caption{(Color online) Estimates of $\eta/s$ using both transport and hydrodynamic
approaches by different groups. The results are expressed in terms of multiples of
the KSS limit.}
\label{fig4}
\end{center}
\end{figure}

Fig~\ref{fig4} shows the comparison of the estimated $\eta/s$ from the experimental data
at LHC energies from various approaches to one obtained in the current work. 
The $\eta/s$ value extracted in this work is obtained mainly from centrality dependence of $v_{2}$ and $v_{4}$ data. Since the extracted shear viscosity to entropy density ratio shows a centrality dependence, we obtained a finite range of value of $\eta/s$ of the produced fluid for a given center of mass energy. For the present case the range of $\eta/s$ shown in figure~\ref{fig4} includes contribution from both $v_{2}$ and $v_{4}$.
Most of the results seems to indicate the value of $\eta/s$ $<$ 4 times the KSS limit. Two of the approaches shown in the figure are based on microscopic transport theory. The remaining are based on the macroscopic hydrodynamic approaches.  J. Xu et al.~\cite{Xu:2011fe}, have estimated the specific shear viscosity using A Multi Phase Transport model (AMPT). They have explained the experimental data on charged particle pseudorapidity density per participating nucleon pair, $v_{2}(p_T )$  in 40-50\% centrality, and the centrality dependence of $v_2$ for
Pb-Pb collision at $\sqrt{s_{NN}}$ = 2.76 TeV. S. Plumari et al.~\cite{Plumari:2011re}, have used a parton cascade
based approach with a temperature dependent $\eta/s(T)$ to simulate $v_{2}$ at LHC energies.
They compare the simulated results to experimental measurements of $v_{2}(p_{T})$ for
20-30\% collision centrality Pb-Pb collisions at  $\sqrt{s_{NN}}$ = 2.76 TeV.
P. Bozek~\cite{Bozek:2011wa} has estimated the value of  $\eta/s$ by using a 2+1D viscous hydrodynamic model in
which a Glauber based entropy density initialization
was considered. In addition to shear viscosity a bulk viscosity to entropy density of 0.04 was also considered in the simulation. Further the freeze-out and resonance decay contributions were based on THERMINATOR
event generator. M. Luzum~\cite{Luzum:2010ag} uses a 2+1D viscous hydrodynamic simulations to carry out
shear viscous evolution starting from a smooth initial condition for LHC energies.
B. Schenke et al.~\cite{Schenke:2011tv},
uses a 3+1D viscous hydrodynamics with fluctuating initial conditions to explain the experimental data
on $v_{2}(p_{T})$ and $p_{T}$ integrated $v_{2}$ for various collision centralities to estimate the value of $\eta/s$
at LHC energies.  H. Song~\cite{Song:2012tv} has used a hybrid approach
to explain the measured $p_{T}$ integrated $v_{2}$ as
well as $p_{T}$ differential $v_{2}$ for various collision centrality in Pb-Pb collisions at
$\sqrt{s_{NN}}$ = 2.76 TeV.
The hybrid approach combines the macroscopic viscous hydrodynamic description for the QGP
fluid with the microscopic hadron cascade model for the subsequent evolution of the hadronic stage.

\section{SUMMARY}
In summary, within the framework of Israel-Stewart's 2nd order hydrodynamics, we
have simulated $\sqrt{s_{NN}}$=2.76 TeV Pb-Pb collisions at midrapidity. Two different 
initial conditions based on , (i) Glauber model and (ii) Color Glass Condensate (CGC), are used to model the initial transverse energy density
distribution in Pb-Pb collision.  The simulations are carried out for
$\eta/s$ values between 0.0 to 0.16, using a lattice + hadron resonance
gas model based equation of state which has a cross over temperature for the
quark-hadron transition at 175 MeV. The shear viscous corrections are
considered both in the evolution equations and in the freeze-out distribution
function.

We have compared our simulated results to the experimental data at
midrapidity on the centrality dependence  of $dN_{ch}/d\eta$, $v_{2}$, and $v_{4}$  as
a function of $p_{T}$ of charged hadrons measured in Pb-Pb  collisions at
$\sqrt{s_{\rm {NN}}}$ = 2.76 TeV.  The comparisons have been done for the
collision centrality classes of 10-20\%, 20-30\%, 30-40\%, and 40-50\%.
We find there is nice agreement between the experimental measured $dN_{ch}/d\eta$
and the simulated results for both Glauber and CGC initial conditions.
The $v_{2}(p_{T})$ experimental data requires a slightly lower value of $\eta/s$
for simulations with Glauber model initialization compared  to the CGC based initial conditions.
For a Glauber based initial condition the data prefers a $\eta/s$ value between 0.0 to 0.12,
while for CGC based initial condition it prefers a value between 0.08 to 0.16.

The experimental data for $v_{4}(p_T )$ in 0-10\% centrality collision measured with respect to
 $\Psi_4$ is underestimated for ideal and shear viscous fluid evolution in both Glauber and CGC initialization. For other collision centralities considered in this study the experimental 
 $v_4(\Psi_{4})$ as well as  $v_4(\Psi_{2})$ data prefers a shear viscosity to entropy density ratio between 0 (for 20-30\% centrality) - 0.12 (for 40-50\% centrality) for both Glauber and CGC initial condition. Like elliptic flow, the experimental data for $v_{4}(p_{T})$  demands higher values of shear viscosity for more peripheral collisions. However, we want to point out that the quality of the agreement between simulation and experimental data for centrality dependence of $v_{4}(p_{T})$ is inferior compared to $v_{2}$. A better description of $v_{4}(p_{T})$ could be possible through an event-by-event hydrodynamics simulation in which $v_{4}(p_{T})$ is proportional to the combination of both second and fourth order eccentricities of the initial state~\cite{Teaney:2012ke}.

The current work uses smooth initial conditions. A more realistic approach
would be to use a fluctuating initial condition and carry out event-by-event
hydrodynamics~\cite{Schenke:2011bn,Qiu:2011iv,Gardim:2011xv,Holopainen:2010gz}. 
This will enable us to study the odd flow harmonics $v_{3}$
along with the even harmonics $v_{2}$ and $v_{4}$. We have already started working to incorporate the fluctuating initial condition in our 2+1D viscous hydrodynamics code. Our recent study for ideal hydrodynamics with fluctuating initial condition can be found in \cite{RihanHaque:2012wp}. The simultaneous description of all these experimentally measured flow harmonics in a
viscous hydrodynamics framework will probably provide a better
estimation of $\eta/s$ at the LHC energies. We plan to consider
some of these effects in the near future. The current results however
agrees well with other estimates of $\eta/s$ for the same experimental data.
These estimates are based on both transport and hydrodynamic approaches.
All results seems to indicate that the value of $\eta/s$ for the QCD matter
formed Pb-Pb collisions at $\sqrt{s_{NN}}$ = 2.76 TeV is between 1 - 4 $\times$ 1/4$\pi$
(the KSS limit). This value is very close to that obtained for QCD matter at RHIC
energies ($\sqrt{s_{NN}}$ = 200 GeV)~\cite{Roy:2012jb,Niemi:2011ix,Shen:2010uy,Chaudhuri:2009uk,Bozek:2009dw,Schenke:2010rr,Romatschke:2007mq}.


\ack
VR and BM are supported by DAE-BRNS project Sanction No.  2010/21/15-BRNS/2026.

\end{document}